\documentclass[11pt]{article}
\usepackage{geometry}                
\usepackage{graphicx}
\usepackage{amssymb}
\usepackage{epstopdf}
\usepackage[T1]{fontenc}
\usepackage{makeidx}
\usepackage{color}
\usepackage{tabularx}
\usepackage{amsmath}
\usepackage{amsthm}
\usepackage{bbm}
\usepackage{latexsym}
\usepackage[applemac]{inputenc}
\usepackage{pdfsync}
\newcommand{\dt}{\mathfrak{t}}

\newtheorem{defi}{Definition}[section]
\newtheorem{prop}[defi]{Proposition}
\newtheorem{theorem}[defi]{Theorem}
\newtheorem{conj}[defi]{Conjecture}
\newcommand{\beconj}{\begin{conj}}
\newcommand{\enconj}{\end{conj}}
\newcommand{\betheo}{\begin{theorem}}
\newcommand{\entheo}{\end{theorem}}

\newcommand{\beprop}{\begin{prop}}
\newcommand{\enprop}{\end{prop}}

\newcommand{\prf}{\noindent{\bf Proof.\,\,\,}}

\newcommand{\C}{\mathbb C}

\newcommand{\N}{\mathbb N}

\newcommand{\R}{\mathbb R}

\def\deq{\stackrel{\mathrm{def}}{=}}

\title{On a generalization of the binomial distribution and its Poisson-like limit}

\author{
E.M.F.  Curado$^{\mathrm{a,b,c}}$,  J. P. Gazeau$^{\mathrm{b}}$, 
 Ligia M.C.S. Rodrigues$^{\mathrm{a}}$(\footnote{e-mail:
 evaldo@cbpf.br,
gazeau@apc.univ-paris7.fr, ligia@cbpf.br} )\\
\emph{$^{\mathrm{a}}$ Centro Brasileiro de Pesquisas Fisicas,}\\
\emph{Rua Xavier Sigaud 150, 22290-180 - Rio de Janeiro, RJ, Brazil}\\
\emph{$^{\mathrm{b}}$ Laboratoire APC,
Universit\'e Paris  Diderot,}   \\
\emph{10, rue A. Domon  et  L. Duquet
75205 Paris Cedex 13, France} \\
\emph{$^{\mathrm{c}}$ Instituto Nacional de Ciencia e Tecnologia}}

\date{\today}                                           

\begin{document}
\maketitle

\begin{abstract}
We examine a generalization of the binomial distribution associated with a strictly increasing sequence of numbers and we prove its Poisson-like limit. Such generalizations might be found in quantum optics with imperfect detection. We discuss under which conditions this distribution can have a probabilistic interpretation.
\end{abstract}

\section{Introduction}
\label{intro}

In most of the realistic models in Physics one must take correlations into account .  Physical correlations, if large enough,  could imply statistical correlations. Events which are usually presented as independent, like in  a binomial Bernoulli process, are actually submitted to correlative perturbations, as small as they  could be. These perturbations or fluctuations with  respect to the independent case lead to deformations of the mathematical independent laws.  Examples are found in all areas of Physics. As a matter of fact, the deformation of the Poisson distribution upon which is based the construction of Glauber coherent states in quantum optics leads to the so-called nonlinear  coherent states  (see \cite{dodonov02} and references therein). Also, the realization of a special class of these states, corresponding
to a special choice of deformation, has been proposed in the quantized motion of a trapped atom in a Paul trap \cite{matosvogel96,kivodav01}.

 In a recent work \cite{cugaro10},  we have  examined  the possibilities of using in quantum measurement a certain class of such nonlinear or non-poissonian coherent states defined as  a perturbation of the standard poissonian coherent states.
 In the case of  imperfect detection, we  have subsequently deformed the underlying Bernoulli distribution, and  this raises interesting and non-trivial questions on the statistical content of such sequences. These questions are examined in  the present paper. 

The organization of this note is as follows. In Section \ref{genbin} we define a Bernoulli-like distribution associated to an arbitrary strictly increasing sequence of positive numbers. This distribution is  ``formal'' because it is not always positive. In  Section \ref{theo} we prove that for a certain class of such sequences the proposed Bernoulli-like distribution has a positive Poisson-like law as its limit when $n \to \infty$. Section \ref{qdef} is devoted to a non trivial example beyond the sequence of non-negative  integers, namely  the sequence of $q$-brackets and their corresponding $q$-binomial or \emph{Gaussian coefficient}. In  Section \ref{posconst} we discuss the question of positiveness in the general case and we examine in Section  \ref{probinter} under which conditions a probabilistic interpretation can be given to the proposed Bernoulli-like distribution. 

 \section{A generalization of the binomial distribution from a sequence of numbers}
\label{genbin}

In a process of a sequence of $n$ trials with two possibles outcomes, ``win'' and ``loss'', the probability of obtaining $k$ wins is given by the binomial or Bernoulli distribution. This can be expressed as
\begin{equation}
\label{Bernoulli}
p_k^{(n)}(\eta) = \binom{n}{k}\eta^k\, (1-\eta)^{n-k},
\end{equation}
where $\eta$ is the probability of having the outcome ``win'' and $1 - \eta$, the outcome ``loss''. Therefore, we can say that to the sequence of non-negative integers  $n \in \N$ there corresponds the Bernoulli binomial distribution above.  As it is well-known, for $\eta = t/n$, in the limit $n \to \infty$ the probability of having $k$ ``wins'' is the Poisson-law, $\exp(-t) t^k/k! $.

Let us now consider  an infinite, strictly increasing countable sequence of nonnegative real numbers $\{ x_n\}_{n\in \N}$. With no loss of generality, we suppose that $x_0 = 0$ and $x_1 = 1$. In the sequel, we will use the symbol $\mathcal{X}$ to designate generically such sequences.

With each sequence $\mathcal{X}$ defined above we can construct a Bernoulli-like distribution,
\begin{equation}
\label{berndef}
\mathfrak{p}_k^{(n)}(\eta) = \frac{x_n!}{x_{n-k}! \, x_k!}\, \eta^k \, p_{n-k}(\eta)\, ,
\end{equation}
where the factorials $x_n!$ are given by
\begin{equation}
\label{factorgen}
x_n! = x_1\, x_2\, \cdots x_n\, ,\quad x_0! \deq 1\, ,
\end{equation}

satisfying
\begin{equation}
\label{sumberndef}
\quad \sum_{k=0}^n \mathfrak{p}_k^{(n)}(\eta)=1.
\end{equation}

Among various possible probabilistic interpretations of the distribution (\ref{berndef}), one seems to be closer to the Bernoulli one.  Let us introduce for $n>0$ the ratio $y_n = x_n/n$ and define $y_0 = 0$.  Then (\ref{berndef}) can be rewritten as
\begin{equation}
\label{berndef2}
\mathfrak{p}_k^{(n)}(\eta) =\binom{n}{k}  \eta^k \, \varpi_{n,k}(\eta)\, ,
\end{equation}
where $\varpi_{n,k}(\eta) \deq \dfrac{y_n!}{y_{n-k}! \, y_k!}\,p_{n-k}(\eta)$. The above expression might be still interpreted as the probability, after $n$ trials, to have $k$ wins and $n-k$ losses, but the probability to get the latter has changed under the effect of some correlation. We will come back to this point in Section \ref{probinter}.

In (\ref{berndef}) the functions $p_{n}(\eta)$ trivially satisfy $p_n(0) = 1$, $p_n(1) = 0$, for all $n$, and the sequence $(p_{n})_{n\in \N}$ is determined from (\ref{sumberndef}) by recurrence:
\begin{align}
\label{recpn}
\nonumber p_0(\eta) &= 1\,, \ p_1(\eta) = 1- \eta\, , \ p_2(\eta) = (1-\eta)\left(1 -\left(\frac{x_2}{x_1} -1\right)\eta\right)\, , \\
\nonumber p_3(\eta) &= (1-\eta)\left(1 -\left(\frac{x_3}{x_1} -1\right)\eta + \left(1 -\frac{x_3}{x_1}\left(\frac{x_2}{x_1} -2\right)\right)\eta^2\right) \, , \cdots\\
p_n(\eta) &= 1 - \eta^n - \sum_{k=1}^{n-1}\frac{x_n!}{x_{n-k}! \, x_k!}\, \eta^{n-k} \, p_{k}(\eta)\, .
\end{align}

With the sequence $\mathcal{X}$ is associated the generalized ``exponential'' $\mathcal{N}(\dt)$  according to
\begin{equation}
\label{expongen}
\mathcal{N}(\dt) = \sum_{n=0}^{\infty} \frac{\dt^n}{x_n!}\, .
\end{equation}
Since the sequence $\mathcal{X}$ is strictly increasing, the limit of $x_n$ as $n \to \infty$ exists and  the d'Alembert's Ratio Test shows that it is equal to  the radius of convergence, say $R_{\mathcal{X}}$,   of $\mathcal{N}(\dt) $:  $R_{\mathcal{X}} = \lim_{n\to \infty} x_n \leq \infty$.

The Bernoulli-like distribution (\ref{berndef}) has a remarkable property.
\beprop
\label{ntnetat}
Let $\mathcal{N}(\dt)$ be the generalized exponential (\ref{expongen}) associated to a sequence $\mathcal{X}$. Let $\eta_m \geq 0$ the number defined as follows: 
\begin{equation}
\label{etam1}
\mbox{for all}\ \eta \in [0,\eta_m]\, ,   \, p_n(\eta) \geq 0\ \forall n\in \N\, . 
\end{equation}
Then the following identity holds true  for all $\eta\in [0,\eta_m]$: 
\begin{equation}
\label{prop}
\sum_{s=0}^{\infty} p_s(\eta) \frac{\dt^s}{\mathcal{N}(\dt)\,x_s!} = \frac{1}{\mathcal{N}(\eta\dt)}\, .
\end{equation}
\enprop
\prf 
Since $p_n(0) = 1$ for all $n$, two cases are to be considered: 
\subparagraph{$\eta_m = 0$} The identity (\ref{prop}) is trivially verified since $\mathcal{N}(0)=1$.
\subparagraph{$\eta_m > 0$}
The identity (\ref{prop}) comes straight from the definition of $\mathcal{N}(\dt)$. Starting from 
\begin{equation}
\label{prop2}
\mathcal{N}(\eta\dt)\sum_{s=0}^{\infty} p_s(\eta) \frac{\dt^s}{\,x_s!} = \mathcal{N}(\dt)\, ,
\end{equation}
and writing $\mathcal{N}(\eta\dt)$ explicitly we have
\begin{equation}
\label{prop3}
\sum_{j=0}^{\infty}\frac{\eta^j\dt^j}{x_j!}\sum_{s=0}^{\infty} p_s(\eta) \frac{\dt^s}{\,x_s!} = \mathcal{N}(\dt)\, . 
\end{equation}
Since for all $\eta \in [0,\eta_m]$ we have $0\leq p_n(\eta) = \mathfrak{p}_0^{(n)}(\eta) \leq 1$ for all $n$, the fact that all terms of the  double series (\ref{prop3}) are nonnegative allows to invert the order of summation:
by a simple change of variable, $s +j = n$, we get
\begin{equation}
\label{prop4}
\sum_{n=0}^{\infty}\dt^n\sum_{s=0}^{\infty} p_{n-s}(\eta) \frac{\eta^s}{x_s!x_{n-s}!} = \mathcal{N}(\dt)\, ; 
\end{equation}
from property (\ref{sumberndef}) we have that 
\begin{equation}
\sum_{s=0}^{\infty} p_{n-s}(\eta) \frac{\eta^s}{x_s!x_{n-s}!} = \frac{1}{x_n!}\, 
\end{equation}
and therefore the left hand side of relation (\ref{prop4}) above is exactly the definition of $\mathcal{N}(\dt)$. 
\qed

Another useful result concerns the general expression of the polynomial $p_n(\eta)$.
\beprop
\label{invexp}
Polynomials $p_n(\eta)$ have the alternative expression:
\begin{equation}
\label{alterpn}
p_n(\eta) = \sum_{k=0}^n (-1)^k\,\binom{x_n}{x_k}\, I_k \, \eta^k\, , 
\end{equation}
where 
\begin{equation}
\binom{x_n}{x_k} = \frac{x_n!}{x_k! x_{n-k}!}
\end{equation}
and the numbers $I_k$ are defined  through the expression of the inverse of the positive series (\ref{expongen}) with $\dt \geq 0$:
\begin{equation}
\label{invexp}
\lbrack \mathcal{N}(\dt) \rbrack^{-1} = \sum_{n=0}^{\infty} (-1)^n\,  I_n\, \frac{\dt^n}{x_n!}\, .
\end{equation}
This definition, at least formally,  becomes effective for all $\dt $
smaller than the convergence radius of the series (\ref{invexp}).
\enprop
\prf 
From the identity $\lbrack \mathcal{N}(\dt) \rbrack^{-1} \, \mathcal{N}(\dt)  = 1$ we derive the following recurrence relation for the coefficients $I_k$:
\begin{equation}
\label{recIk}
(-1)^n \, I_n = \sum_{k=0}^{n-1}(-1)^{k-1}\, I_k\, \binom{x_n}{x_k}\, , \quad I_0 = 1\, , \quad I_1 = 1\,. 
\end{equation}
Let us inject expression (\ref{alterpn})  into relation (\ref{recpn}):
\begin{align*}
p_n(\eta) &= 1 - \eta^n - \sum_{k'=1}^{n-1} \frac{x_n!}{x_{n-k'}! \, x_{k'}!}\, \eta^{n-k'} p_k(\eta) \\ 
&= 1 - \eta^n - \sum_{k'=1}^{n-1} \frac{x_n!}{x_{n-k'}! \, x_{k'}!}\, \eta^{k'}\sum_{k''=0}^{n-k'} (-1)^{k''}\frac{x_{n-k'}!}{x_{n-k'-k''}! \, x_{k''}!}\, I_{k''}\, \eta^{k''}  \, ;
\end{align*}
after the change $(k',k'') \mapsto (k = k'+k'', k'') $ on the summation indices, we have
\begin{align*}
p_n(\eta) &= 1 - \eta^n - \sum_{k=1}^{n-1} \frac{x_n!}{x_{n-k}! \, x_{k}!}\, \eta^{k}\sum_{k''=0}^{k} (-1)^{k''}\frac{x_{k}!}{x_{k-k''}! \, x_{k''}!}\, I_{k''} \\
&= \sum_{k=0}^n (-1)^k\,\binom{x_n}{x_k}\, I_k \, \eta^k \, .
\end{align*}
Then, equating the coefficients of $\eta^k$ on both expressions for $p_n(\eta)$ above we get
\begin{equation}
\quad I_0 = 1\, , \quad I_1 = 1\,  , \quad  (-1)^n \, I_n = \sum_{k=0}^{n-1}(-1)^{k-1}\, I_k\, \binom{x_n}{x_k}\,  ,
\end{equation}
which is exactly the recurrence relation (\ref{recIk}) for the coefficients $ I_k$. Thus, (\ref{alterpn}) is proven to hold true.
\qed

\section{A limit theorem}
\label{theo}

We now suppose that our sequence $\mathcal{X} = \{ x_n\}_{n\in \N}$ is such that $x_n \to\infty$ as $n \to \infty$. Then  the radius $R_{\mathcal{X}}$ of convergence of the series $\mathcal{N}(\dt)$ is $\infty$. Furthermore, suppose that, at fixed finite $m$, 
\begin{equation}
\label{hypS}
\underset{n \to \infty}{\lim}{\frac{x_{n-m}}{x_n}} = 1\, ,
\end{equation}
 or equivalently, $\underset{n \to \infty}{\lim}{\frac{\Delta_{m\,n}}{x_n}} = 0$, where $\Delta_{m\,n} \equiv x_n - x_{n-m}  $ is the ``distance'' between those two elements of the sequence. We denote by $S$ the class of such sequences, $S = \{\mathcal{X}\,  |\,  x_n \in \mathcal{X} \to \infty\,\,\,    \mbox{as}\,\,\, n \to \infty\}$.

\betheo
\label{poisson1}
For any  sequence $\mathcal{X}$ in $S$ such that the radius of convergence $R_{\mathcal{X}}$ of the series (\ref{invexp}) is not zero, the limit when  $n \to \infty$ of the Bernoulli-like distribution $ \mathfrak{p}_k^{(n)}(\eta)$ in (\ref{berndef}) with  $\eta = \dt/x_n$ is equal to
 a Poisson-like  law:
\begin{equation}
\label{poisslim}
 \mathfrak{p}_k^{(n)} (\eta) = \frac{x_n!}{x_{n-k}! \, x_k!}\, \eta^k \, p_{n-k}(\eta)
 \underset{n \to \infty}{\to} \frac{1}{\mathcal{N}(\dt)}\, \frac{\dt^k}{x_k!}\, .
\end{equation}
\entheo
\prf
Let us rewrite (\ref{alterpn}) as 
\begin{equation}
\label{pntx}
p_n\left(\frac{\dt}{x_n}\right) = \sum_{k=0}^n (-1)^k\, \left(\prod_{l=0}^{k-1}\frac{x_{n-l}}{x_n}\right)\, I_k \, \frac{\dt^k}{x_k!}\, . 
\end{equation}
As $n\to \infty$, ${\lim}{\frac{x_{n-l}}{x_n}} = 1$ and (\ref{pntx}) goes to expression (\ref{invexp}) for $\lbrack \mathcal{N}(\dt) \rbrack^{-1}$ in the interval of convergence of the latter. On the other hand, the factor $\frac{x_n!}{x_{n-k}! \, x_k!}\, \eta^k$ in (\ref{poisslim}) with $\eta = \dt/x_n$, can be rewritten as
\begin{equation*}
\prod_{l=0}^{k-1}\frac{x_{n-l}}{x_n}\,\frac{\dt^k}{x_k!}\, ;
\end{equation*}
due to the same assumption (\ref{hypS}), this factor goes to $\dt^k/x_k!$ as $n\to \infty$.

\qed

An  important example of sequences in the class $\mathcal{S}$ are the 
Delone sequences \cite{alietal09}, which are infinite strictly increasing sequences of
nonnegative real numbers
\begin{equation}
\label{seq}
\mathcal{X} = \{ x_n\}_{n\in \N}\, , \qquad x_0 = 0\, ,
\end{equation}
 with  the following two constraints :
\begin{itemize}
  \item[(d1)] $\mathcal{X}$ is \emph{uniformly discrete} on the positive real line $\R^+\, : $ $\exists r>0$ such that $x_{n+1} - x_n \geq r$ for all $n\in N$, which means that there exists a minimal distance between two successive elements of the sequence,
  \item[(d2)] $\mathcal{X}$ is \emph{relatively dense} on $\R^+\, :$ $\exists R>0$ such that for all $x \in \R^+$ $\exists n \in \N$ such that $\vert x - x_n\vert < R$, which means that there exists a maximal distance, say $L$,  between two successive elements of the sequence.
\end{itemize}
These conditions imply that  $\lim_{n \to \infty} x_n = \infty$ and $\Delta_{m\,n}/n \leq L\,m/n \to 0$ as $n\to \infty$. 

It should also be noticed
that from this point of view sequences like
$x_n = n^{\alpha}\,( \log{n})^\beta, \, \alpha > 0$ and any $\beta$ or like $x_n = (\log{n})^\beta$, $\beta > 0$ or, even more generally, $x_n = n^\alpha(\log(\log(...){n})^\beta$ are included in $S$.  

On the other hand,  this is not the case for familiar deformations of integers like
$q$-brackets \cite{koorn},
\begin{equation}
\label{deform}
[n]_{q} = \frac{1-q^n}{1-q}\,  ,
\end{equation}  
nor in general for sequences $\mathcal{X}$ with $x_n$ increasing exponentially with $n$. 

 \section{Examples from  $q$-calculus}
\label{qdef}

So far we did not question whether or not the sequence of Bernoulli-like distributions (\ref{Bernoulli}) defines true probability distributions, i.e. if they are non-negative within the range $\eta \in [0,1]$.  An illuminating example enjoying  such a probabilistic interpretation is precisely yielded by the above mentioned $q$-brackets. Let us consider the sequence $\mathcal{X}_q$ of these  $q$-deformations of non-negative integers:
\begin{equation}
\label{qbrack}
x_n = [n]_{q} \deq \frac{1-q^n}{1-q} = 1 + q + \dots + q^{n-1}\, , \quad q >0\, . 
\end{equation}
This sequence  is strictly increasing with $x_0= 0$ and $x_1= 1$. With the notation
\begin{equation}
\label{qcalcul1}
(a;q)_k \deq \prod_{l=0}^{k-1}(1-aq^l)\,, 
\end{equation}
the factorial $x_n!$ reads as
\begin{equation}
\label{qfact1}
x_n! = \frac{(q;q)_n}{(1-q)^n}\, . 
\end{equation}
In this case, the generalized binomial coefficients 
\begin{equation}
\label{qbin}
\binom{x_n}{x_k} = \prod_{l=0}^{k-1}\frac{1- q^{n-l}}{1-q^{l+1}}\equiv \binom{n}{k}_q
\end{equation}
bear the name of Gaussian coefficients (\cite{koorn}). The polynomials $p_n(\eta)$ simply factorize as:
\begin{equation}
\label{factorgauss}
p_n(\eta) = \prod_{l=0}^{n-1}\left( 1 - q^l\, \eta\right)\, , 
\end{equation} 
and expand as
\begin{equation}
\label{expandgauss}
p_n(\eta) = \sum_{k=0}^n (-1)^k\,\binom{n}{k}_q\,  q^{\dfrac{k(k-1)}{2}}\, \eta^k\,.
\end{equation}
We note that $I_k = q^{\frac{k(k-1)}{2}}$. Also note  how the corresponding generalized exponential $\mathcal{N}(\dt) \equiv \mathcal{N}_q(\dt)  $ is related  to the $q$-exponentials of the $q$-calculus \cite{koorn}:
\begin{equation}
\label{Nqexp}
\mathcal{N}_q(\dt) = e_q( (1-q)\dt) \, , \quad \mbox{for} \quad q <1\, ,  
\end{equation}
where
\begin{equation}
\label{qexp}
e_q(z) \deq  \sum_{n=0}^\infty \frac{z^n}{(q;q)_n} = \frac{1}{(z;q)_{\infty}}\,,  \quad q<1\, ,\quad \vert z\vert < 1\, , 
\end{equation}
and 
\begin{equation}
\label{Nqexp}
\mathcal{N}_q(\dt) =  E_{q^{-1}} \left(\left(1-q^{-1}\right)\,\dt\right)\, , \quad \mbox{for} \quad q > 1\, ,  
\end{equation}
where
\begin{equation}
\label{qexp}
E_q(z) \deq  \sum_{n=0}^\infty \frac{q^{\frac{n(n-1}{2}}z^n}{(q;q)_n} = (-z;q)_{\infty}\,,  \quad q<1\, , \quad z \in \C\, . 
\end{equation}
In these formulas, $(a;q)_{\infty}\deq \prod_{k=0}^{\infty}(1-aq^k)$. Also note that   the inverse of the generalized exponential is explicit here, due to the relation $e_q(z)\,E_q(-z)= 1$:
\begin{equation}
\label{qinvgen1}
\frac{1}{\mathcal{N}_q(\dt) } = E_q( -(1-q)\dt)  \, ,\quad \mbox{for} \quad q <1\, , 
\end{equation}
and
\begin{equation}
\label{qinvgen1}
\frac{1}{\mathcal{N}_q(\dt) } = e_{q^{-1}} \left(-\left(1-q^{-1}\right)\,\dt\right)\, , \quad \mbox{for} \quad q >1\, . 
\end{equation}

\begin{figure}
\begin{center}
\includegraphics[width=3in]{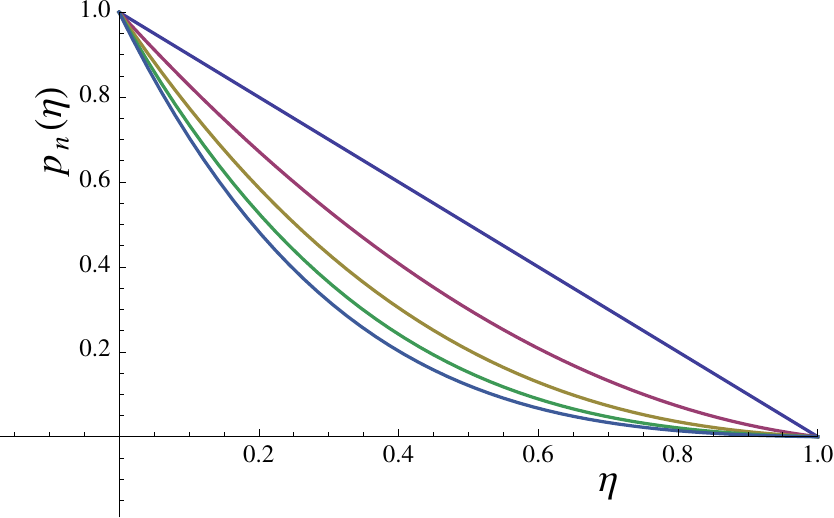}
\caption{Behavior of $p_n(\eta)$, Eq. (\ref{expandgauss}), for $n = 1, 2, 3, 4, 5$  and $q=4/5$. 
$p_n(\eta)$ is a true probability for $\eta \in [0,1]$.
The curves correspond to increasing $n$ from top to bottom. }
\label{fig1}
\end{center}
\end{figure}

\begin{figure}
\begin{center}
\includegraphics[width=3in]{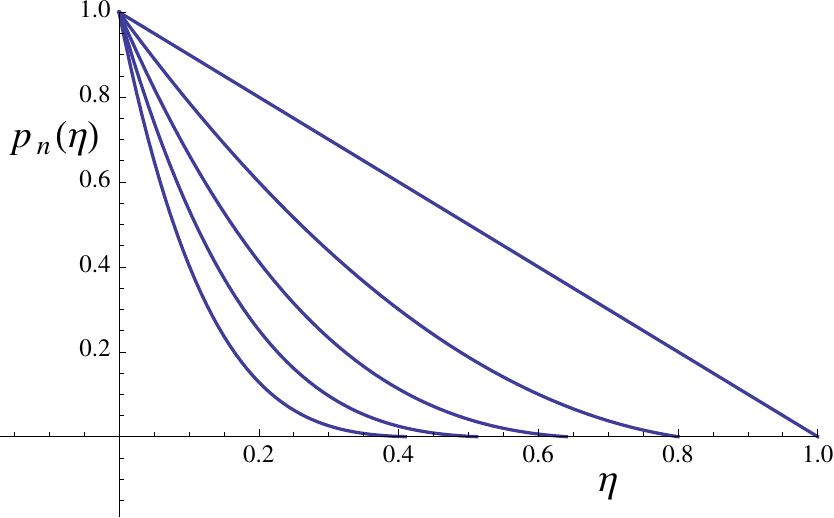}
\caption{Behavior of $p_n(\eta)$, Eq. (\ref{expandgauss}), for $n = 1, 2, 3, 4, 5$  and $q=5/4$. Only the positive part of $p_n(\eta)$ is shown, that is, the curves are cut at each $\eta = (5/4)^{1-n}$. They correspond to increasing $n$ from top to bottom. }
\label{fig2}
\end{center}
\end{figure}

\begin{figure}
\begin{center}
\includegraphics[width=3in]{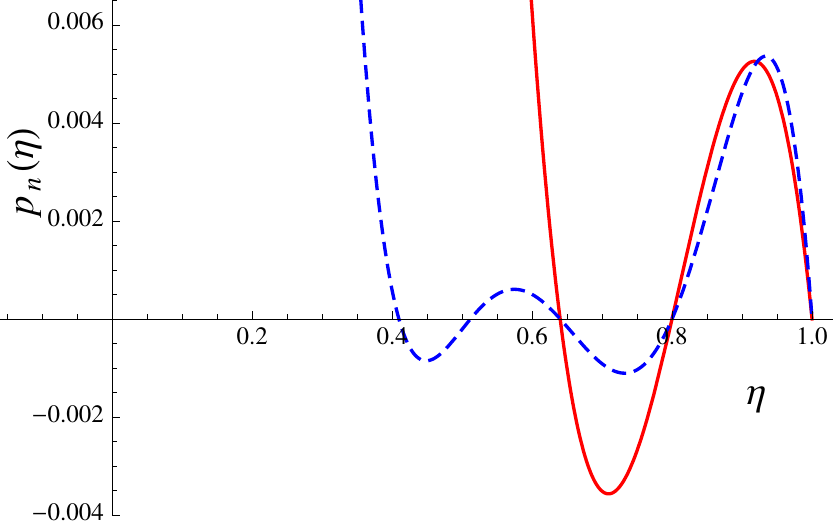}
\caption{Behavior of $p_3(\eta)$  (full line) and $p_5(\eta)$ (dashed line) for $q=5/4$, showing their negative 
parts. }
\label{fig3}
\end{center}
\end{figure}

Hence we have here to consider two cases.
\begin{itemize}
  \item[(i)]  If $q< 1$, the sequence $\mathcal{X}_q$ is bounded: $x_n \to 1/(1-q)$ as $n \to \infty$ and expression (\ref{berndef}) is a true probability distribution in the whole range $\eta \in [0,1]$. We then have for the generalized exponential the finite convergence radius $R_{\mathcal{X}_q} = 1/(1-q)$. In Fig. (\ref{fig1}) we show the behavior of $p_n(\eta)$ as given by 
 (\ref{expandgauss}) for a few values of $n$ and $q=4/5$. 
  \item[(ii)] If $q> 1$, the sequence $\mathcal{X}_q$ is unbounded: $x_n \to \infty$ as $n \to \infty$. At a given $n$, the expression (\ref{berndef}) is a true probability distribution in the restricted range $\eta \in [0,q^{1-n}]$. The corresponding generalized exponential has an infinite convergence radius $R_{\mathcal{X}_q} = \infty$. In Fig. (\ref{fig2}) we show the behavior of $p_n(\eta)$ as given by (\ref{expandgauss}) for a few values of $n$ and $q=5/4$. The largest 
  values of $\eta$ for which $p_n(\eta)$ is a true probability are respectively $(5/4)^{1-n}$. In 
  Fig. {\ref{fig3}}, as an example,  we show the behavior of $p_3(\eta)$ and $p_5(\eta)$ for values of $\eta$ larger than $(4/5)^2$ and $(4/5)^4$, respectively. Note that the oscillations increase with growing values of $n$.    
\end{itemize}

It is interesting to observe that such $q$-deformations of non-negative integers are quite rigid within the present context. As a matter of fact, let us explore the possibility of obtaining another  non-trivial example of sequence $\mathcal{X}$ starting from an infinite  sequence of positive numbers $a_0=1, \, a_1, \dotsc , a_n,\, \dotsc$ such that the corresponding sequence of polynomials $p_n(\eta)$ factorizes as 
\begin{equation}
\label{seqqgen}
p_n(\eta) = \prod_{l=0}^{n-1}(1-a_{l}\, \eta)\, ,
\end{equation}
which means that they obey the simple recurrence relation
\begin{equation}
\label{recqgen}
p_n(\eta) = (1-a_{n-1}\, \eta)\,  p_{n-1}(\eta)\, , \quad p_0(\eta) = 1\, ,
\end{equation}
which has to be consistent with recurrence (\ref{recpn}).
\begin{prop}
The only sequence $a_0=1, \, a_1, \dotsc , a_n,\, \dotsc$ which makes compatible  (\ref{recpn}) and (\ref{recqgen}) is the power sequence $a_n = q^n$, where $q=a_1$, and  for which the corresponding sequence $\mathcal{X}$ is precisely  $\mathcal{X}_q$.
\end{prop}
\prf
The proof is obtained by recurrence. It is obviously true for $n=1$, since $p_1(\eta) = (1-\eta)$ obeys (\ref{recpn}). Suppose that the assertion is true for all $k\leq n-1$, i.e. $p_{k}(\eta) = \prod_{l=0}^{k-1}(1-a_{l}\, \eta)$ and  $x_l = [l]_q$ for all $k,\,l \leq n-1$. Let us put $\eta = q^{-1}$ in (\ref{recpn}). Since $p_k(q^{-1}) = 0$ for all $2\leq k \leq n$, Equation (\ref{recpn}) gives:
\begin{equation*}
0= 1 - q^{-n} - x_n\,q^{1-n}\, p_1(q^{-1}) = 1 - q^{-n} - x_n\,q^{1-n}\,(1-q^{-1})\, , 
\end{equation*}
and this leads to $x_n = [n]_q$. Inserting this expression of $x_n$ into Equation (\ref{recpn}) and using the recurrence assumption again gives the expression (\ref{seqqgen}) for $p_n$. 

\qed

\section{The positiveness constraint for probabilistic interpretation}
\label{posconst}

We now examine the question of positiveness of the distribution $\mathfrak{p}_k^{(n)}(\eta)$ or equivalently of the polynomials $p_n(\eta)$. Let us divide the set $\Sigma$ of our sequences $\mathcal{X}$ into two subsets $\Sigma_{\pm}$, according to whether or not  the corresponding  polynomials  $p_n(\eta)$ remain non-negative  in the interval $[0,1]$ for all $n$:
\begin{equation}
\label{posdef}
\Sigma_{+} = \{ \mathcal{X}\, | \, \forall\,n \geq 0\, , p_n(\eta) \geq 0 \ \mbox{for all}\ \eta \in [0,1]\}\,, \quad \Sigma_{-} = \Sigma\setminus\Sigma_{+} \, .
\end{equation} 
Sets  $\Sigma_{\pm}$ are not empty since $q$-calculus provides non-trivial examples for both. Any sequence $\mathcal{X}$ in $\Sigma_{+}$ has a probabilistic interpretation as a generalization of the binomial law. To date we have been unable to give an analytical example  in $\Sigma_{+}$ which is not a sequence of $q$-brackets. 

For a sequence $\mathcal{X}$ in $\Sigma_{-}$, each corresponding polynomial $p_n(\eta)$ has at least one root in the open interval $(0,1)$. Let us designate by $1/a_{n-1}$, $a_{n-1}\geq 1$ the root of $p_n(\eta)$ which is the closest one to 0. Since $p_n(0)=1$ for all $n$, it is clear that $p_n(\eta) >0$ for all $\eta \in [0,1/a_{n-1})$. We are thus in presence of a  sequence of numbers $a_0=1, \, a_1, \dotsc , a_n,\, \dotsc$, all $> 1$,  such that for $n > 1$, with $\rho_0 = 1$, the  sequence of polynomials $p_n(\eta)$ factorizes as
\begin{equation}
\label{factpn}
p_n(\eta) = (1-\eta)\,(1-a_{n-1}\,\eta)\,\rho_{n-2}(\eta)\, , \quad \rho_s(\eta) = 1 + \sum_{k=1}^{s}(-1)^k b_k^{(s)} \,\eta^{k}\, , 
\end{equation}
where the polynomial $\rho_s(\eta)$ is supposed to be positive in $\eta \in [0,1/a_{n-1})$. Comparing the $n$ orders of $\eta$ in expressions (\ref{factpn}) and (\ref{alterpn}) we have $n$ equations, thus allowing us to express the $n$ numbers $x_n,\, b_1, \dotsc, b_k,\dotsc, b_{n-2}, I_n$ in terms of the $n-1$ numbers $a_1,\, a_2, \dotsc, a_{n-1}$. 
We first determine by recurrence the numbers $x_n$ and $I_n$ in terms of $a_{n-1}$ and of the previous values of $x_k$, $I_l$ for $1\leq k,l \leq n-1$.
\begin{prop}
\label{xnfctan}
\begin{itemize}
  \item[(i)] The $n^{\mathrm{th}}$ element $x_n$ of the sequence $\mathcal{X}$ is given in terms of $a_{n-1}$ and of $x_k$, $I_l$ for $1\leq k,l \leq n-1$ by:
\begin{equation}
\label{xnan}
x_n =  \frac{[n]_{a_{n-1}}}{\sum_{k=1}^{n-1}(-1)^{k-1}\dfrac{x_{n-1}!}{x_{n-k}!\, x_k!}\,I_k\,[n-k]_{a_{n-1}}}\, . 
\end{equation}
  \item[(ii)] Similarly, the $(n+1)^{\mathrm{th}}$ coefficient  $(-1)^n\, I_n$ of the polynomial $p_n(\eta)$ in its form (\ref{alterpn}) is given by:
\begin{equation}
\label{Inan}
(-1)^n\, I_n = \frac{\sum_{k=1}^{n-1}(-1)^{k}\dfrac{x_{n-1}!}{x_{n-k}!\, x_k!}\,I_k\,a_{n-1}^{n-k}\, [k]_{a_{n-1}}}{\sum_{k=1}^{n-1}(-1)^{k}\dfrac{x_{n-1}!}{x_{n-k}!\, x_k!}\,I_k\,[n-k]_{a_{n-1}}}\, . 
\end{equation}
\end{itemize}
We recall that $I_1 = 1$ and $[n]_q = (1 - q^n)/(1 - q)$. The case $n=2$, for which $\rho_0(\eta) = 1$  gives immediately the relations
\begin{equation}
\label{neq2}
x_2 = 1+ a_1=[2]_{a_1}\, , \quad I_2 = a_1\, . 
\end{equation}
\end{prop}
\prf
The proof is straightforward. It is enough to use the fact that $\eta = 1$ and $\eta = a_{n-1}^{-1}$ are roots of $p_n(\eta)$:
\begin{align*}
0= \, &p_n(1) = \sum_{k=0}^{n} (-1)^k\, \binom{x_n}{x_k}\, I_k\\
0= \, &a_{n-1}^{n} \, p_n(a_{n-1}^{-1}) = \sum_{k=0}^{n} (-1)^k\, \binom{x_n}{x_k}\, I_k\, a_{n-1}^{n-k}\, . 
\end{align*}
Taking the difference between these two expressions gives (\ref{xnan}). Now, from the first one we have the recurrence relation for $I_n$,
\begin{equation*}
(-1)^n\, I_n = -1 - x_n\,\sum_{k=1}^{n-1} (-1)^k\,\dfrac{x_{n-1}!}{x_{n-k}!\, x_k!} \, I_k\, .
\end{equation*}
The expression (\ref{Inan}) is then obtained by substituting (\ref{xnan}) in the equation above and using the fact that $[n]_q- [n-k]_q= q^{n-k}\, [k]_q$.
\qed

These relations, when applied recursively, allow to express the numbers $x_n$ and $I_n$, and so the polynomial $p_n$, uniquely in terms of the sequence $a_1,\, a_2,\, \dotsc, a_{n-1}$.

We have now the general result which allows to determine recursively all coefficients of the polynomial $\rho_{n-2}(\eta)$.

From the identification of the coefficients of the powers of $\eta$ in  the two alternative expressions (\ref{factpn}) and (\ref{alterpn}) of $p_n(\eta)$, we can construct the linear system 
\begin{equation}
\label{linearvn}
\mathrm{M}_n\,v_n = {}^t(x_{n-1}\,(1+a_{n-1}),\,x_2\,a_{n-1}, 0, \dotsc,\,0)\, \,  ,
\end{equation}
where $v_n$ is the row vector $v_n= {}^t(x_n,\, b_1, \dotsc, b_k, \dotsc, b_{n-2},\,I_n)$ and $M_n$ is the $n \times n$ matrix below:

$$
\left(
\begin{array}
[c]{ccccc}%
I_1 & -1& 0 & 0 & \dotso \\
\frac{x_{n-1}}{x_2} I_2& -1-a_{n-1} &  -1& 0 &\dotso \\
\frac{x_{n-1}x_{n-2}}{x_3 x_2} I_3& -a_{n-1}&  -1-a_{n-1} &  -1 & 0  \\
\frac{x_{n-1}!}{x_4! x_{n-4}!} I_4& 0 & -a_{n-1}& -1-a_{n-1} & \dotso \\
\vdots& \dotso & \ddots & \ddots & 0 \\
\frac{x_{n-1}!}{x_k! x_{n-k}!} I_k & 0 & \dotso&  0 & 0\\
\vdots & \dotso & \ddots & 0 & 0\\
I_{n-1} & 0 & \dotso &  0 & 0\\
0 & 0 & \dotso & 0 & 0
\end{array}
\right. $$
$$\left.
\begin{array}
[c]{ccccc}%
0&\dotso  &\dotso&\dotso&0\\
 0&\dotso&\dotso&\dotso&0\\
0& \dotso  &\dotso&\dotso&0\\
 &&&&\vdots\\
-1-a_{n-1} &-1&0&&\vdots\\
\vdots & \dotso & \ddots &&\vdots\\
0  &\dotso&  -a_{n-1}&-1-a_{n-1}&0\\
 0 & \dotso& & -a_{n-1}&1
\end{array}
\right)\, .
$$
For the $k$-th line of $M_n {^t}v_n$, $1\leq k\leq n-1$, we have:
\begin{equation*}
\frac{x_{n-1}!}{x_k! x_{n-k}!} I_k x_n - b_{k-2} a_{n-1} - b_{k-1} (1+a_{n-1}) - b_k \,\,  ,
\end{equation*}
with $b_{-1}= b_0 = b_{n-1}= b_n=0$; the $n$-th line is 
\begin{equation*}
-a_{n-1} b_{n-2} + I_n \,\, .
\end{equation*}

The determinant of the matrix $M_n$ is given by
\begin{equation}
\label{detMn}
\det{\mathrm{M}_n} =  \sum_{k=1}^{n-1}(-1)^{n+k-1}\dfrac{x_{n-1}!}{x_{n-k}!\, x_k!}\,I_k\,[n-k]_{a_{n-1}} \,\, .
\end{equation}
As a corollary, we can show that the $n^{\mathrm{th}}$ element $x_n$ of the sequence $\mathcal{X}$ and $I_n$ are given by:
\begin{align}
\label{xnan1}
\nonumber x_n = \, &(-1)^n \frac{[n]_{a_{n-1}}}{\det{\mathrm{M}_n}}\\
I_n = \, &\frac{\sum_{k=1}^{n-1}  (-1)^ {k-1}  \frac{x_{n-1}!}{x_{n-k}! x_k!}  I_k a^ {n-k}_{n-1} [k]_{a_{n-1}}}{\det{\mathrm{M}_n}}\,\, .
\end{align}

Since only the  two first components of the vector  $v_n$ are  non-zero,  the numerator of the expression of $x_n$ involves the two minor determinants $\mathrm{MIN}_{11}^{(n)}$ and $\mathrm{MIN}_{21}^{(n)}$ only.  From elementary linear algebra, we have
\begin{equation*}
x_n = \frac{1}{\det{\mathrm{M}_n}}\left(\mathrm{MIN}_{11}^{(n)} (1+a_{n-1}) - \mathrm{MIN}_{21}^{(n)}  a_{n-1}\right)\, .
\end{equation*}
Identifying this expression with (\ref{xnan}) and 
using the relation $[n]_q = (1+q) [n-1]_q -q [n-2]_q$ we can see 
that $ \mathrm{MIN}_{k1}^{(n)}$ is proportional to $[n-k]_{a_{n-1}}$. 
A verification of the values of the determinant for the cases $n=2, 3, 4, 5$ and $6$ 
gives the expression of the minors $\mathrm{MIN}_{k1} $:
\begin{equation*}
\mathrm{MIN}_{k1}^{(n)}  = (-1)^n \,[n-k]_{a_{n-1}}\, . 
\end{equation*}
The determinant of $\mathrm{M}_n$ can thus be written as in Eq. (\ref{detMn}) and the expression (\ref{xnan1}) for $x_n$  follows.

\qed

\begin{figure}
\begin{center}
\includegraphics[width=3in]{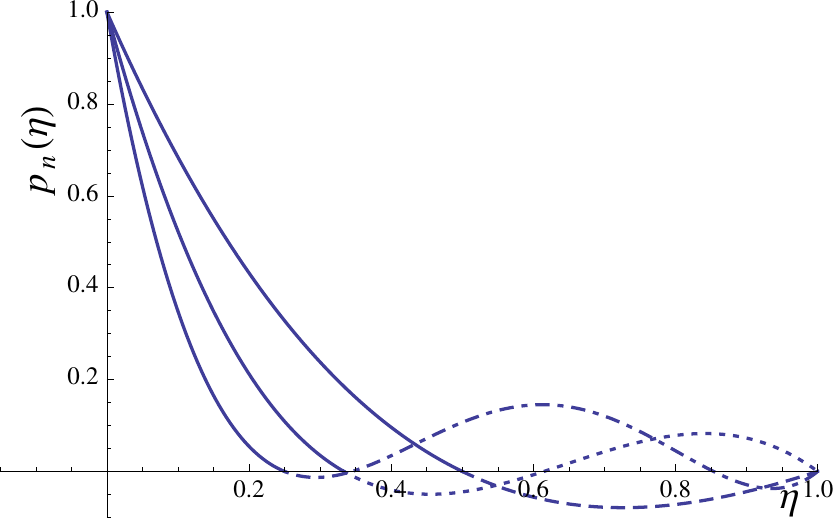}
\caption{Behavior of $p_n(\eta)$ for the sequence $a_n =n$ and $n =  2, 3, 4 $. The continuous 
curves correspond to increasing $n$ from top to bottom.}
\label{fig4}
\end{center}
\end{figure}

As an example, we have examined the case where the sequence of roots of $p_n$ which are closest to $0$ is given by $a_n = n$ . In Fig. {\ref{fig4}} we see the behavior of $p_n$ for $n =  2, 3, 4$.

Now we have to play a delicate recursive game in order to control the consistency of our virtual choice as which $a_{n-1}^{-1}$ is precisely the smallest positive root of the polynomial $p_n$. By ``virtual'', we mean that given a sequence $\mathcal{X}$ in $\Sigma_-$, we know that, by definition, there exists at each step $n$ such smallest positive root $a_{n-1}^{-1}$ which is a function of the set $x_1,\, x_2,\, \dotsc, \, x_n$: $a_{n-1} = A^{\mathrm{min}}_{n}(x_1,\, x_2,\, \dotsc, \, x_n)$. What Proposition  \ref{xnfctan} states is  the reciprocal of these relations (for each step $n = 1,\, 2,\,3,\, \dotsc$): $x_{n} = X_{n}(a_1,\, a_2,\, \dotsc, \, a_{n-1})$, which is interesting by itself as a direct generalization of the $q$-bracket case where $a_{n-1} =q^{n-1}$. 
But the formula (\ref{xnan}) is valid \emph{a priori} for \underline{any} root of $p_n$  which is different from $1$. Determining the function $A^{\mathrm{min}}_{n}$ is not an easy task!

Conversely, starting from a sequence $\mathcal{A}= (a_1,\, a_2,\, \dotsc, \, a_{n-1},\, \dotsc)$ of numbers $ \in (1, \infty)$, 
Proposition  \ref{xnfctan}  yields a deterministic procedure to find a sequence $\mathcal{X}=(x_0=0, x_1=1,\, x_2,\, \dotsc, \, x_n,\, \dotsc)$. It is as well a not easy task to check it, but we can conjecture that
\begin{itemize}
  \item[(i)]  $\mathcal{X}$ is a strictly increasing sequence of positive numbers, 
  \item[(ii)] At each step $n$, the polynomial $\rho_{n-2}$ has no root in $(0,a_{n-1}^{-1})$.
\end{itemize}

\section{A probabilistic interpretation}
\label{probinter}

Let us finally  analyze under which conditions our generalized Bernoulli distribution can be given a probabilistic interpretation.

In a process of $n$ trials ruled by the Bernoulli distribution (\ref{Bernoulli}), 
i. e., independent trials, 
let us consider the case $n=1$. We have two different independent possibilities, 1 win or 1 loss, which are given by
\begin{itemize}
\item $p_1(\eta) = \eta$ - the probability of 1 win;
\item $p_0(\eta) = (1 - \eta)$ - the probability of 1 loss.
\end{itemize}
If we now consider $n=2$, we have 3 different independent possibilities, respectively given by
\begin{itemize}
\item $p_2 (\eta)= \eta^2$ - the probability of 2 wins;
\item $p_1(\eta) = 2 \eta (1 - \eta)$ - the probability of 1 win and 1 loss, independently of the order of the events;
\item $p_0(\eta) = (1 - \eta)^2$ - the probability of 2 losses.
\end{itemize}

In a similar process where the distribution is our Bernoulli-like, Eq. (\ref{berndef}), a simple calculation for $n=1$ gives us 
\begin{align*}
\mathfrak{p}_1^{(1)} (\eta) &= \eta\,,   \mathfrak{p}_0^{(1)} (\eta) = (1-\eta)\, .
\end{align*}
Although this is the same result as for the Bernoulli distribution, Eq. (\ref{Bernoulli}), and we could be led to interpret $\mathfrak{p}_k^{(n)}$  as the probability to have  $k$ wins 
and $n-k$ losses in $n$ trials, when we examine the result for $n=2$, we obtain

\begin{align*}
\mathfrak{p}_2^{(2)} (\eta) &= \eta^2\,, \ \mathfrak{p}_1^{(2)} (\eta) = \frac{x_2}{x_1}\eta(1- \eta)\, , \ \mathfrak{p}_0^{(2)} (\eta) = (1-\eta)[1+(1-\frac{x_2}{x_1})\eta]\, . 
\end{align*}
\noindent
Examining the expressions above we see that the meaning of $\mathfrak{p}_k^{(n)}$ is not obvious at all, because the non-negativity of $\mathfrak{p}_1^{(2)}$ and $\mathfrak{p}_0^{(2)}$ depends on the values of $\eta$ and on the chosen sequence $\mathcal{X}$. For $n=3$, the same happens for both $\mathfrak{p}_0^{(3)}$, $\mathfrak{p}_1^{(3)}$, 
and $\mathfrak{p}_2^{(3) }$ which are, respectively
\begin{itemize}
\item $\mathfrak{p}_0^{(3) } (\eta) = 1 - \frac{x_3}{x_1} \eta (1-\eta) [1 + (1-\frac{x_2}{x_1})] + \frac{x_3}{x_1} \eta^2 (1-\eta) + \eta^3$
\item $\mathfrak{p}_1^{(3) } (\eta) = \frac{x_3}{x_1} \eta (1-\eta) [1 + (1-\frac{x_2}{x_1})\eta]$ 
\item $\mathfrak{p}_2^{(3) } (\eta) = \frac{x_3}{x_1} \eta^2 (1-\eta) $ .
\end{itemize}

Therefore it will be necessary to analyse the non-negativity of $\mathfrak{p}_k^{(n)}$ in function of $\eta$ and the parameters appearing in each sequence $\mathcal{X}$ in order to know under which conditions we can give its associated Bernoulli-like distribution a probabilistic interpretation. As we have seen in Section $4$ when 
$x_n = \frac{1-q^ n}{1 - q}$ and $q < 1$ this positivity problem does not exist and all the 
$p_n(\eta)$ are  non-negative for $0 \leq \eta \leq 1$. The same does not happen for 
$q > 1$; in this case, the non-negativity of $p_n(\eta)$ is assured only in the restricted range $0 \leq \eta \leq q^{1 - n}$. 

In fact, two scenarios are possible in which the $p_k(\eta)$ are non-negative, a "condicio sine qua non" in order to have a probabilistic interpretation. 

In the first, for each "$n$" there is an upper 
limit for $\eta$ ($\eta_{max}^{(n)} \leq 1$), in such a way that $p_m(\eta)$ becomes negative for $m>n$ and $\eta > \eta_{max}^{(n)} $. This limit decreases with "$n$". 

In the second scenario, $0 \leq \eta \leq 1$, all $\mathfrak{p}_k^{(n)}$ satisfy the Bernoulli scheme. The  resulting cost is that the $x_n$'s cannot reach $n$, and have 
increasing differences with $n$ as it increases. 

The first scenario, in which  the $x_n$'s can become larger than $n$ but 
the allowed $\eta$'s become smaller and smaller for increasing values of $n$, was 
shown in Figs. $2$, $3$ and $4$. 
We have shown that in this case we can find sequences $\mathcal{X} = \{ x_m\}_{m\in \N}$ such that the $p_n(\eta)$ are positive for $\eta \leq\ \eta_{max}^{(n)}$. For such sequences, we can interpret $\mathfrak{p}_k^{(n)}(\eta)$  as a probability but we must also have a consistent interpretation for 
$\eta_{max}^{(n)}$.  Two examples were mentioned in the Sections 4 and 5: the case of $q$-calculus for $q > 1$, for which the behavior of $p_n(\eta)$ is shown in Figs. 2 and 3  and the case shown in Fig. 4. 

The second scenario, in which $\mathfrak{p}_k^{(n)} (\eta)$, for all $k$ and $n$,  
 are non-negative for $0 \leq \eta \leq 1$ can be seen in the case of $q$-calculus for 
 $q <1$, see Fig. 1.

Now that we can assure the non-negativity of 
$p_n(\eta)$, and consequently of $\mathfrak{p}_k^{(n)} (\eta)$, $k \leq n$, 
we can analyze in more detail the interpretation sketched at the beginning of 
section \ref{genbin}. We will do this by using as example the $q$-calculus, discussed   
in section \ref{qdef}.

\begin{figure}
\begin{center}
\includegraphics[width=2in]{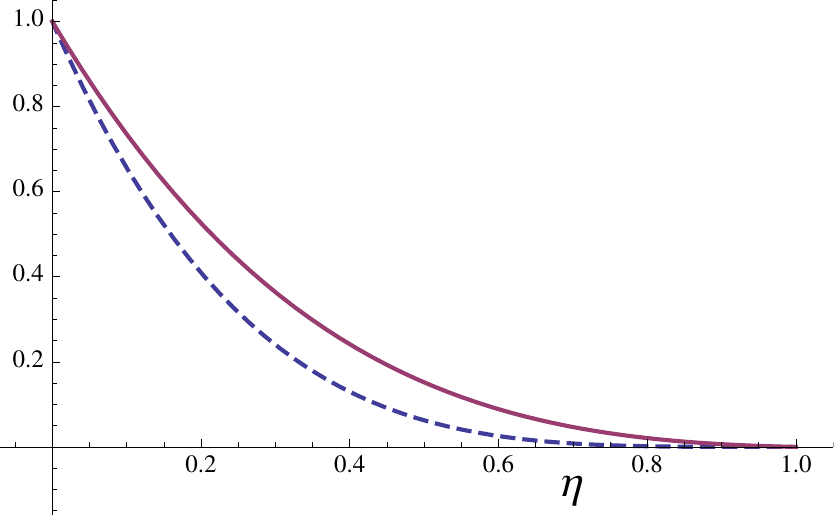}
\caption{$\varpi_{4,0} (\eta)$ (solid line) and $(1-\eta)^4$ (dashed line) in function of $\eta$ for $q = 4/5$. }
\label{pql1}
\end{center}
\end{figure}

\begin{figure}
\begin{center}
\includegraphics[width=2in]{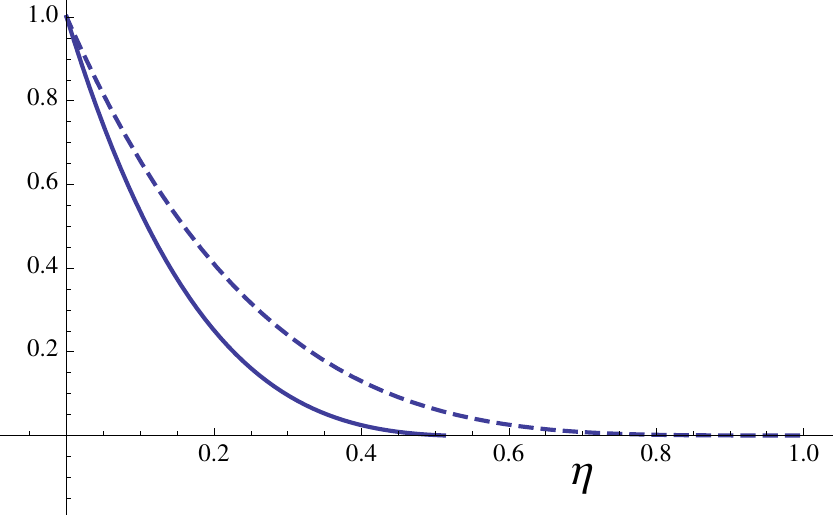}
\caption{$\varpi_{4,0} (\eta)$ (solid line) and $(1-\eta)^4$ (dashed lline) in function of $\eta$ for $q = 5/4$}
\label{pqg1}
\end{center}
\end{figure}

The expression for  $p_n(\eta)$ in this case is shown in Eq.(\ref{factorgauss}). Adopting the probabilistic interpretation suggested by Eq.(\ref{berndef2}), the deformed probability to have $n$ losses in $n$ trials, $ \varpi_{n,0} (\eta)$, 
see Eq. (\ref{berndef2}),  
is always equal to $p_n(\eta)$.  Comparing  $\varpi_{n,0} (\eta)$ (or, equivalently, $p_n(\eta)$) with the Bernoulli 
term associated with to get $n$ losses in $n$ trials, $(1-\eta)^n$, 
we can immediately see that, for $q < 1$, $\varpi_{n,0} (\eta)$ is always greater than the Bernoulli non disturbed term $(1-\eta)^n$. This means that in this case, 
$q<1$, we have correlations that increase the probability to get repeated losses, while the probability to get repeated wins remain unchanged. The mixed cases, with 
$k$ wins and $n-k$ losses have decreasing probabilities at the expense of the increasing of the term  $\varpi_{n,0} (\eta)$. 
The other way round occurs for $q>1$, where the probability to get losses decreases with respect to the Bernoulli non disturbed case. Both cases are shown in Figs. \ref{pql1} and \ref{pqg1} for $n=4$. These are the typical behaviors for the other values of $n$. 

We would like to comment that we are considering here only a class of 
correlations, where the probability to get repeated losses increases (or decreases) with respect to 
the standard Bernoulli case. Certainly there are many other kinds of correlations that 
are not included in the Bernoulli deformations we have presented and that lead 
to other kinds of perturbation not considered here.

\end{document}